**Spatially resolving superconductivity in type-II superconductors**

Donald M. Evans[1,2,3]*, Michele Conroy[4], Lukas Puntigam[1], Dorina Croitori[5], Lilian Prodan[1], Marin Alexe[2], James O. Douglas[4], Baptiste Gault[4,6], and Vladimir Tsurkan[1,5]

[1] Experimental Physics V, Center for Electronic Correlations and Magnetism, Institute of Physics, University of Augsburg, Augsburg 86159, Germany.

[2] Department of Physics, University of Warwick, Coventry CV4 7AL, UK.

[3] Department of Sustainable Energy Technology, SINTEF Industry, Oslo, Norway

[4] Department of Materials, Imperial College London, London SW7 2AZ, UK.

[5] Institute of Applied Physics, Moldova State University, Chisinau, Moldova.

[6] Max-Planck-Institut für Eisenforschung GmbH, Max-Planck-Straße 1, Düsseldorf 40237, Germany.

*E-mail: donald.evans@sintef.no

**Abstract:** Superconductivity is identified by the emergence of a macroscopic zero-resistance state, typically inferred from a vanishing four-probe voltage at finite current. That inference assumes spatially uniform conduction—e.g., at least one continuous superconducting path between the current leads and voltage electrodes that sample a finite potential gradient—and can fail if the drive current bypasses the electrodes or if narrow filaments short the current contacts. Here we introduce a methodology to test these assumptions in superconductors, by using spatially resolved measurements of local variations in dc using cryogenic conductive atomic-force microscopy (cAFM). Using Fe(Se,Te) as a model system, we find that despite bulk measurements consistent with a homogeneous superconducting state, the material exhibits a heterogeneous conducting landscape: micrometre-scale superconducting regions coexist with relatively insulating areas. We further show that cAFM resolves conductance fluctuations at 20 K (> $T_C$) that vary between repeated scans, consistent with expectations for short-lived, pre-formed Cooper pairs in the BCS–BEC crossover regime. These results establish cAFM as a practical tool to validate assumptions underlying four-probe transport and underscore the need for direct spatial probes in materials whose macroscopic response can conceal nanoscale inhomogeneity. Accurate identification of macroscopic properties is critical for materials classes like superconductors that are defined by their macroscopic properties.

**Introduction** Type-II superconductors are widely studied for their technological potential and for the rich physics emerging from their correlated electron behaviour.[1–5] Like all superconductors, they are defined by their macroscopic transport signatures: a sharp transition to zero DC resistivity and the onset of the Meissner effect (complete magnetic field expulsion).

However, neither of these measurements inherently reveals the spatial extent or volume fraction of the superconducting phase.[6,7] Consequently, they cannot directly determine whether the superconducting response is spatially uniform or arises from a minority phase—for example, via percolation. Examples of zero-potential drop in clearly heterogeneous superconductors are the alkali metal intercalate iron selenides, and arrays of Nb dots.[8–11]

To measure resistivity in superconductors, four-probe methods—such as the van der Pauw technique—are routinely used, as they eliminate contact resistance and allow for precise determination of zero-resistance states.[12] However, these techniques rely on an essential assumption: that the sample is electronically homogeneous.[6,12] When this condition is not met, the current distribution within the material becomes complex and unpredictable, breaking the link between the measured voltage and intrinsic resistivity.[6] In such cases, the measured resistance reflects not an intrinsic material property, but a geometry- and connectivity-dependent response that cannot be reliably interpreted. Other powerful techniques involve using microprobe but these also require assumptions about uniform current path.[13]

Techniques such as point-contact spectroscopy (PCS) provide localised electronic measurements with a well-established theoretical basis.[14] In the conventional "needle-anvil" geometry, a sharp metallic tip forms a contact with the superconductor—typically tens of microns across.[15] When the contact size is smaller than the electron mean free path, the system enters the ballistic (Sharvin) regime, where the differential conductance (dI/dV) reflects the local quasiparticle excitation spectrum. PCS can thus probe the superconducting gap amplitude and symmetry at specific sites.[15,16] Insightful as it is, it does not provide spatially resolved maps of local DC conductivity variations, the prerequisite for validating four-probe transport measurements.

In other fields, conductive atomic force microscopy (cAFM) is an established technique for directly mapping local variations in conductivity. A nanoscale conductive tip is rastered across the surface under applied bias, simultaneously recording current and topography to generate spatially resolved 3D maps. Originally developed for characterising semiconducting and insulating surfaces, cAFM operates in a geometry analogous to the "needle-anvil" configuration used in PCS (see Supplementary Note 1 and Figure S1).[17] It is routinely applied in ferroelectrics and Mott insulators, with cryogenic implementations functioning down to ~1.5 K.[18,19] Crucially, cAFM delivers co-registered maps of conductivity and topography over tens of microns with sub-5 nm lateral resolution—directly resolving the macroscopic length scales relevant to four-probe transport. In addition to imaging, cAFM can apply local electric fields or mechanically modify the surface, enabling controlled studies of termination, dead layers, and buried features. [20–23] These combined capabilities make cAFM uniquely suited for testing the assumptions. We note that there are multiple other techniques regularly applied to superconductors that allow conductivity to be evaluated, rather than directly measured, representative examples are discussed in Supplementary Note 2.

In this work, we demonstrate cryogenic conductive atomic force microscopy (cAFM) on a superconductor for the first time. Using Fe(Se$_{0.4\pm\alpha}$Te$_{0.6\pm\beta}$) as a model system, we show that while bulk measurements suggest a homogeneous superconducting state, cAFM reveals a strikingly heterogeneous conductivity landscape. By correlating the local conductivity with topographic features we are able to extract site specific needles for atom probe tomography (APT), we find that the majority insulating phase corresponds to the previously accepted superconducting composition, while superconductivity localises in regions closer to FeSeTe$_2$. Beyond identifying this misassignment, cAFM detects conductivity fluctuations above $T_s$ = 14.4 K—consistent with preformed Cooper pairs, long predicted but not previously directly observed in this system. These results establish cAFM as a powerful tool for superconductivity: (i) validating assumptions underlying four-probe transport, (ii) directly linking electronic, structural, and compositional heterogeneity, and (iii) robustly detecting subtle, spatially localised phenomena inaccessible to bulk methods.

**Material background**

Fe(Se$_{0.4\pm\alpha}$,Te$_{0.6\pm\beta}$) exhibits a superconducting transition at $T_C \approx$ 14.5 K, and is known to host inherent atomic-scale disorder due to Se and Te ions occupying the same Wyckoff sites but at different vertical positions.[24–26] It has been reported to display multiband superconductivity, low and spatially inhomogeneous superfluid density, and exotic phenomena including possible Majorana zero modes, topological superconductivity, and spatially varying in-gap states.[27–36] Theoretically, it is situated near the BCS–BEC crossover regime, suggesting the potential for preformed Cooper pairs—though direct spatially resolved evidence of such fluctuations remains absent.[24,25,37,38]

**Apparent homogeneity in bulk measurements**

The bulk properties of our Fe(Se$_{0.4\pm\alpha}$Te$_{0.6\pm\beta}$) crystals exhibit signatures typically associated with homogeneous superconductivity. Figure 1**a** presents the resistivity of Fe(Se$_{0.4\pm\alpha}$Te$_{0.6\pm\beta}$) as a function of temperature. Above the superconducting transition the sample follows metallic like conductivity; at $T_{onset}$ ~14.4 K $\rho(T)$ collapses to zero within $\Delta T$ ~1 K. The narrow transition width and a normal state residual resistivity $\rho_0$ ~300 $\mu\Omega$.cm, among the lowest reported for Fe(Se,Te), are often used as evidence of high quality single crystals devoid of excess Fe.[39] The magnetisation measurements corroborate this picture (Figure 1**b**). In the zero-field-cooled (ZFC) run the volume susceptibility reaches –1 (after demagnetisation correction), signalling nearly ideal flux expulsion. The much smaller field-cooled response is typical of strong vortex pinning rather than incomplete shielding. Taken together, the sharp resistive transition, low $\rho_0$, and full diamagnetic screening are consistent with bulk, spatially uniform superconductivity.

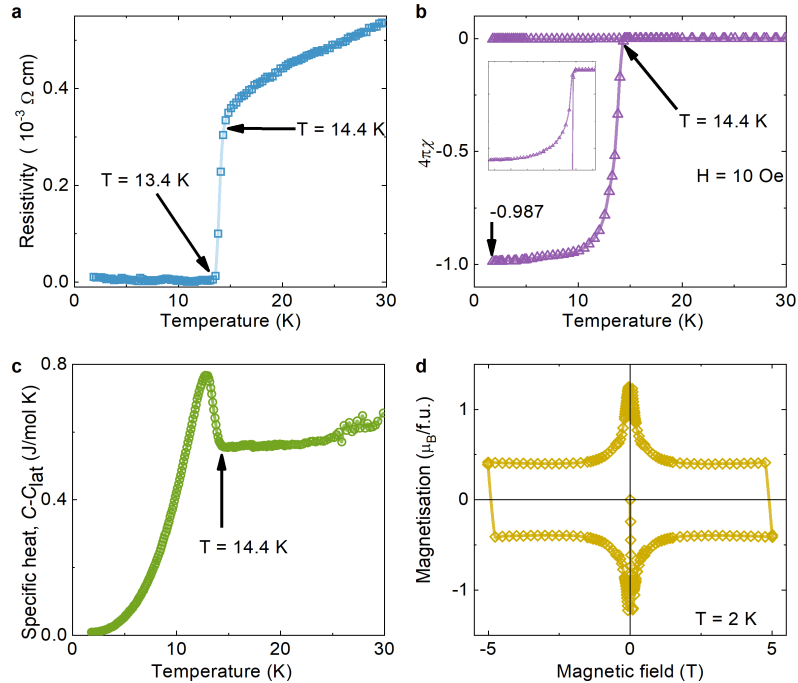

**Figure 1 | Bulk superconductivity below ca. 14 K. a,** Electrical resistivity $\rho(T)$ measured on warming in zero field shows a sharp drop to zero at $T_{onset} \sim 14.4$ K. **b,** Magnetic susceptibility $\chi(T)$ recorded under $\mu_0 H = 10$ Oe in zero-field-cooled (ZFC) and field-cooled (FC) modes; the ZFC signal tends towards –1 at 2 K (after demagnetisation correction), indicating nearly complete diamagnetic screening. **c,** Electronic specific heat after subtraction of the lattice contribution, displaying a λ-type anomaly with onset at $T_C = 14.4$ K. **d,** Magnetisation versus field at 2 K shows the symmetric hysteresis loop characteristic of a type-II superconductor.

Figure 1c gives the specific heat data, displaying a sharp λ-type anomaly at on $T_C \approx 14.4$ K. A Debye fit to C/T between 3 K and 5 K (Supplementary Note 3, Figure S2) yields a residual Sommerfeld coefficient $\gamma_0 = 0.807$ mJ / mol K$^2$ —among the lowest reported for this family.[26,39] Calorimetry on crystals from the same growth run puts the electronic superconducting volume fraction at 95–96 % and gives a gap value $\Delta_0 = 2.4(1)$ meV, in good agreement with others in the literature.[26] Crucially, low-temperature specific heat samples only the metallic regions of the crystal: the electronic term scales with the density of states at $E_F$, so semiconducting or insulating areas contribute negligibly. The large jump therefore shows that 95–96 % of the electrons at $E_F$ condense into Cooper pairs, but it does not reveal what fraction of the physical volume is superconducting.

Figure 1**d** gives a magnetic hysteresis loop at 2K. It is symmetric, has an open shape expected for a strong type-II superconductor; applying the Bean model for a thin slab gives a critical current density 1x10$^5$ A/cm$^2$.[40,41] Complementary scanning electron microscopy (SEM) images and energy dispersive x-ray (EDX) maps (Figure S3) show no compositional or morphological heterogeneity within the detection limit of the area imaged in Figures 2-5. Standard interpretation of these bulk probes suggests a high-quality, macroscopically uniform superconductor.

**Spatially resolved measurements**

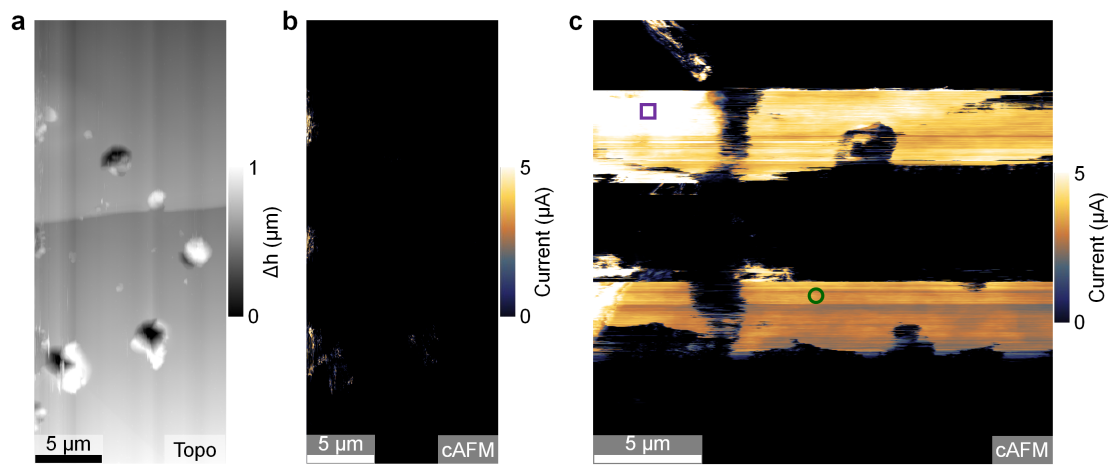

**Figure 2 | cAFM reveals nanoscale electronic inhomogeneity at 1.5 K. a,** Topography of a freshly cleaved Fe(Se$_{0.4 \pm \alpha}$Te$_{0.6 \pm \beta}$) surface (horizontal slow-scan; 15.5 nm pixel); changes in height information is given by the false colours according to the scale bar. **b,** Simultaneous current map acquired with a 100 mV tip–sample bias. The measured current is indicated by the false colour, with yellow pixels giving high currents and black pixels showing no measurable current. **c,** Current map of an adjacent area containing two conducting stripes measured with a vertical slow-scan direction and 17 nm pixel size. Coloured shapes highlight pixels carrying ≈ 5.5 µA (purple square) and ≈ 3.5 µA (green circle).

To visualise the spatial distribution of electrical conduction, we performed conductive atomic-force microscopy (cAFM) at 1.5 K with a tip bias of 100 mV. Figure 2**a** shows the topography of a freshly cleaved surface, transferred into vacuum within < 6 min of cleaving; apart from shallow terraces and a few cleavage pits the surface is smooth. Figure 2**b** is the corresponding and simultaneously collected current map: is the simultaneously acquired current map: over most of the scan the current stays below the measurement noise floor (I ≲ 0.1 nA), interrupted only by sub-micron islands that carry currents in the microampere range.

Figure 2c images an adjacent area that contains two wider stripes of enhanced conductivity. These stripes conduct roughly five orders of magnitude more current than the surrounding regions. Line profiles (Figure S4) show that the current drops from several microamperes to below the noise floor within < 50 nm, demonstrating that the conductive paths are sharply confined; the same abrupt transition is observed for all conducting regions, in orthogonal scan direction. The stripes themselves are also non-uniform: representative pixels carry I ≈ 5.5 µA (purple square) or I ≈ 3.5 µA (green circle), implying at least two distinct local conductance states. We highlight that insulating surfaces with small areas with enhanced conductive were consistently observed across multiple months of measurements on multiple different samples: and the majority of scanned areas showed no current at the measured voltage. For clarity the topography, intensity, and cAFM for the forward and backward scans of Figure 2**c** are presented in Figure S5. Critically, the cAFM data of Figure 2 demonstrate that these apparently homogeneous bulk superconducting samples have heterogeneous conductivity.

**Temperature dependent behaviour**

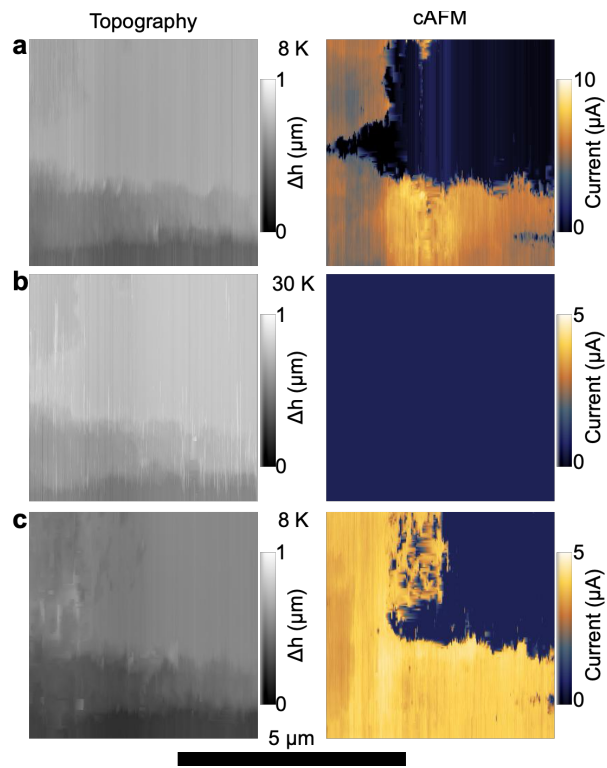

**Figure 3 | Enhanced conductivity above and below T$_C$. a,** Topography and cAFM current map (tip-sample bias = 100 mV) recorded at 8 K; micro-ampere currents mark highly conducting regions. **b,** Topography and cAFM of the same area imaged at 30 K with the same bias: the current falls below the noise floor while the topography is unchanged within the scan resolution. Both the measured topography and the instrument's interferometric stage position confirm this is the same area to within a few pixels. **c**, Topography and cAFM

current map (tip-sample bias = 50 mV) of the same area after re-cooling to 8 K; the re-emergence of high currents over largely the same area confirms their superconducting origin. All scans use a 15 nm pixel size and a horizontal slow-scan axis; grey shades encode height, while the blue–yellow scale encodes current.

To confirm superconductivity as the origin of the observed currents, we acquired cAFM maps of the same region above and below the macroscopic superconducting transition temperature. Figure 3**a** shows the topography and cAFM of a selected area at 8 K; micro-ampere currents appear that do not clearly correlate with the topography. Figure 3**b** presents the same area imaged at 30 K. Although the topography is unchanged within experimental error, the current is quenched everywhere, so at these voltages the surface appears insulating—even though there is a good tip–sample contact, as evidenced by topography image. Figure 3**c** displays the topography and cAFM after re-cooling the sample below $T_C$: high conductivity re-emerges, as expected for a superconductor. A slight change in the extent of the superconducting regions is observed after recooling; we speculate that this may stem from measurement-induced changes in local stoichiometry or structure.[20–22] We note that (a) the interferometric laser positioning system, together with the topography, confirms that the same area was imaged in all scans, and (b) the temperature was changed with the tip lifted from the surface; the force set-point was recalibrated once thermal equilibrium was reached to ensure identical tip pressure at each temperature.

Together with the 1.5 K data, these measurements demonstrate that (i) cAFM can resolve superconductivity locally; (ii) nominally homogeneous Fe(Se,Te) crystals host a complex micro-texture of superconducting and non-conducting regions; and (iii) the superconducting regions vanish at 30 K.

**Preformed cooper pairs**

Theory places Fe(Se,Te) in the BCS–BEC crossover, where short-lived Cooper pairs are expected to persist locally above the bulk $T_C$.[37,38,42] While no theory yet predicts how such pairs should appear in a cAFM experiment, one would intuitively look for highly local, metastable conductance that can be perturbed by the scan itself. Figure 4**a** and **b** show forward- and backward-scan images of the same area used in Figure 3, acquired at 20 K. 20 K was chosen as it is well above $T_C$ but below the temperature where all conductivity vanishes. The topography is identical in both scan directions, as expected, yet the cAFM maps differ markedly: patchy regions of enhanced conductivity change their position and density. Many bright spots span a few pixels, confirming continuity between scan lines, but its alteration between forward and backward scans indicates a fragile electronic state that is sensitive to the probing conditions. Interestingly, the fluctuations do not appear in all of the superconducting region, again suggesting a subtle heterogeneity in the superconducting behaviour. This behaviour is consistent with metastable superconducting fluctuations expected in the crossover regime, while artefacts such as tip lift-off or drift are ruled out by the identical topography images.

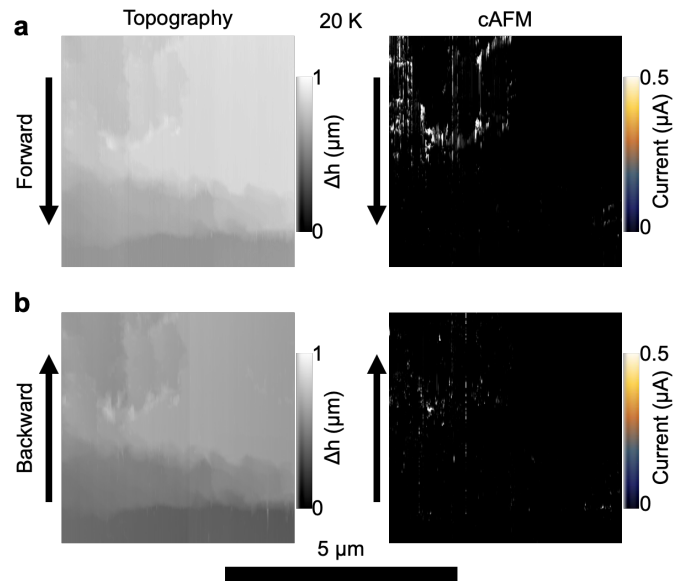

**Figure 4 | Metastable conductivity at $T_C < T < 30$ K. a,** Topography and cAFM current map (tip-sample bias = 50 mV) from the forward scan at 20 K; the bright conducting dots are ≈ 75 nm in diameter (≈ 5 pixels). **b,** Topography and cAFM of the immediately following backward scan. Topography is unchanged, whereas the distribution of conducting dots has altered, confirming the features are metastable. The pixel size is 15 nm with a horizontal slow-scan direction.

**Local composition**

To uncover the origin of the heterogeneous conductivity we turned to atom-probe tomography (APT). The exact cAFM area was relocated in the SEM (Fig. 5a, overlay of SEM and cAFM), and two needle specimens were lifted-out: (1) a non-conducting patch and (2) a neighbouring high-current patch (Methods). Figure 5**b** shows the three-dimensional concentration map for specimen 1. The upper 20 nm are O-rich (blue) and were excluded from the quantitative analysis; a comparable oxide cap was present on specimen 2, consistent with the samples weeks of ambient exposure after the cAFM work.

Figures 5**c** (insulating) and 5**d** (superconducting) compare the summed composition of the two needles. The contrast is unambiguous: the superconducting region is depleted in Fe relative to Te—the Fe/Te ratio is drastically reduced—whereas the nominal target stoichiometry Fe(Se$_{0.4 \pm \alpha}$Te$_{0.6 \pm \beta}$) was found in the needle from the insulating area. While this result would be consistent with bulk reports that excess Fe suppress $T_C$ but the completely different anion ratio is unprecedented.[39] SEM-EDX maps of the same region (Fig. S3) showed no clear

compositional change; while APT is far more sensitive than EDX, additional structural and compositional studies are required to confirm these findings.[43]

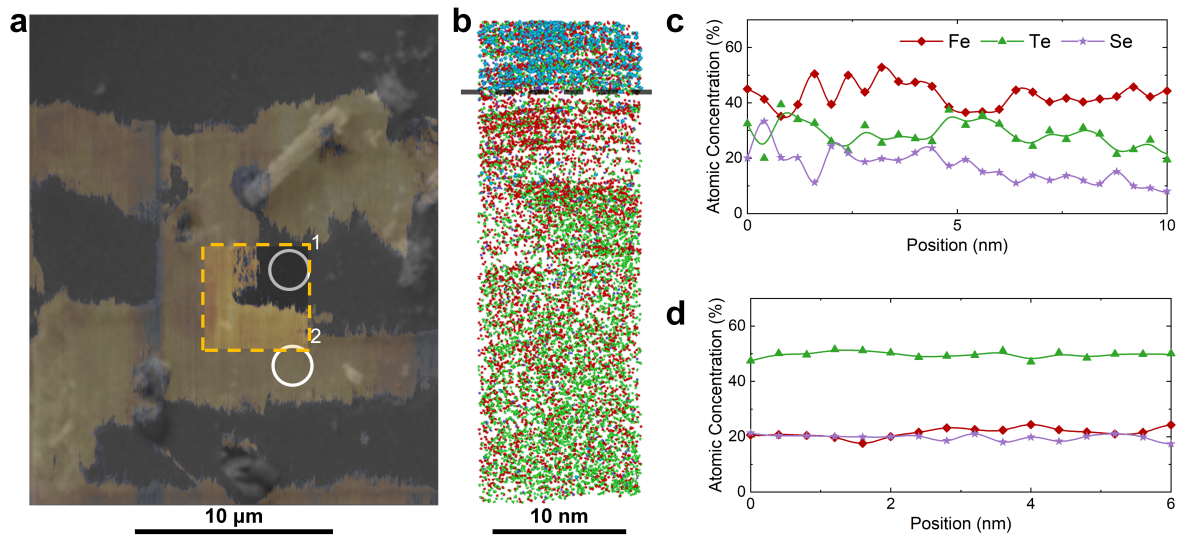

**Figure 5 | Correlating local composition with conductivity. a,** SEM image overlaid with the cAFM current map to locate the region of interest in the FIB-SEM. White circles mark the lift-out sites for atom-probe needles: 1 = insulating region, 2 = conducting region; the yellow rectangle is the cAFM field of view from Figures 3 and 4. **b,** Reconstructed 3-D atom map from needle 1 showing Fe (red), Te (green), Se (purple) and O (blue). The upper ≈ 20 nm O-rich cap (oxide) is omitted from the quantitative analysis. **c,** Depth-integrated concentration profile for the insulating region (needle 1) after oxide removal. **d,** Equivalent profile for the superconducting region (needle 2).

**Implications**

Our results show that a crystal can satisfy the conventional bulk-quality checks—sharp $\rho(T)$ transition, full diamagnetic screening, well-defined specific-heat jump—yet still consist of a heterogeneous superconducting network embedded in an electronically inert matrix. Four-probe transport measurements, which assume a single homogeneous current distribution, are therefore insensitive to such textures and require direct spatial validation before their results are interpreted. By imaging $I(r)$ directly, cAFM provides that missing dimension. Local conductance mapping should thus become a routine companion to macroscopic transport in type-II superconductors, both to verify absolute resistivity values and to guide growth protocols that maximise the volume fraction and connectivity of the superconducting phase. More broadly, any theoretical or experimental analysis of Fe(Se,Te)—and likely other iron-based materials—that relies on bulk averages must allow for the possibility that multiple electronically distinct phases contribute unevenly to the measured signal.

**Conclusions**

We have demonstrated that cryogenic cAFM can deliver spatially resolved DC transport maps of a type-II superconductor, exposing features that bulk probes miss. In Fe(Se,Te) the technique reveals four key findings: (i) superconducting regions are embedded in an insulating matrix; (ii) metastable, spot-like conductance hotspots appear at 20 K (above $T_C$), consistent with pre-formed Cooper pairs in the BCS–BEC crossover; (iii) atom-probe tomography links these conductive regions to a distinct, Fe-depleted local stoichiometry that escapes conventional EDX analysis; and (iv) the complexity of iron-based superconductors may, in part, arise from the coexistence of multiple phases. Together these results show that local transport measurements are indispensable for an accurate experimental description of material properties—an essential first step for optimisation and for developing reliable theoretical frameworks for high temperature superconductivity.

**Experimental methods**

<u>Crystal growth</u>

Single-phase crystals of FeSe$_x$Te$_{1-x}$ were synthesized via a self-flux method using the precursor elements— Fe (99.99% purity), Se (99.999%), and Te (99.999%)—both Se and Te were subjected to zone-refining to minimize oxide contaminants prior to the crystal growth. The synthesis and crystal growth was executed in hermetically-sealed double quartz ampoules. The preparation was done in two-steps. In the first step a mixture of the elements was slowly heated to 1100 °C, soaked for 72 h and then fast cooled to 410 °C and soaked again for 99 h. At the second step, material prepared at the first step was reground in an Ar glove box and filed in quartz ampoule, locked and transferred to pumping system. After pumping to $10^{-3}$ mbar the ampoule was sealed. At the second step, after a fast increase of temperature to 1100 °C, the ampoule was soaked for 72 h followed by slow cooling (6 °C/h) to 410 °C, where final soak for an extended 100-hour period was done. The process was culminating in quenching in a water-ice bath. Elemental stoichiometry was confirmed through EDX, while phase purity was checked on crushed samples via X-ray powder diffraction, employing a STADI-P diffractometer (STOE & CIE) with a position-sensitive detector and CuK$_\alpha$ radiation (λ = 1.540560 Å). For comprehensive analytical details, we refer the reader to previous work by Tsurkan et al. Ref [26] and other characterisations of crystals from the same batch are given by Refs. [36,44,45].

<u>Scanning probe microscopy</u>

The low-temperature scanning probe microscopy (SPM) measurements were conducted with an attoAFM I atomic force microscope (attocube systems AG, Haar, Germany). Boron-doped single-crystal diamond tips (AD-2.8-AS, radius: 10 nm; Bruker France, Wissembourg, France) were employed for these experiments. Prior to measurements, the vacuum chamber was evacuated to pressures below $10^{-5}$ mbar and subsequently purged with helium; this evacuation-purge cycle was executed thrice at ambient conditions. A controlled atmosphere was maintained by introducing helium at a few millibars as an exchange gas. The sample underwent cooling to liquid helium temperatures over a multi-hour period, following our normal protocols.

Sample preparation involved retrieval a virgin crystal from a sealed quartz tube, followed by the fresh cleavage of both crystal facets using a razor blade. One facet was subsequently anchored to an attocube sample holder using a silver conductive paste. The entire preparation procedure—comprising quartz tube breach, crystal cleavage, sample mounting, and chamber evacuation—was expediently completed within an eight-minute timeframe to preclude atmospheric contamination. Several different samples, from the same batch used for bulk measurements, were investigated with the cAFM: all showed a predominantly insulating surface with only local islands of enhanced conductivity.

For all measurements the temperature was changed when the tip was not in contact with the surface. Once the equipment had fully thermalised, for a change in temperature of 20 K this would be about an hour after the target temperature was first reached, the diether (that controls tip pressure on the sample) was retuned to give a consistent pressure between scans.

All the SPM images were analysed and processed using the open source software "Gwyddion" [46] The only processing of the data was changing the false colour, setting the scale, and correcting horizontal scars; the latter being routine for the fast scan direction of an SPM image. All images in Figures 2, 3 and 4 have been cropped, so they fit as display items. None of these made any quantifiable change to raw data and all original (.bcrf) files are available upon reasonable request to the corresponding author.

Bulk data collection

Temperature dependent resistivity was measured on a freshly cleaved sample using the resistivity option of Physical Property Measurements System (PPMS, Quantum Design) with four electrical contacts made using conducting silver paint. The specific heat was measured on the same equipment. The magnetic susceptibility and magnetisation data was measured using a superconducting quantum interface devices (SQUID, MPMS-5, Quantum Design) with a field cooled and zero field cooled protocol. The SQUID measurements where repeated for several different crystals from this batch and no variation in magnetic properties was observed.

Atom probe tomography

Samples were opened in a $N_2$ environment glove-box, mounted onto a copper stub and placed in an atom probe puck (Cameca) for transfer purposes. This was transferred under UHV to Side 1 of a dual puck stage plate (Oxford Atomic) inside a Thermo-Fisher Scientific Helios Hydra 5 CX PFIB using a Ferrovac Vacuum Cryo Vacuum Module for site specific APT sample preparation. A commercially available highly Sb-doped single-crystal silicon micro post array (CAMECA Instruments Inc., Madison, WI, USA) was mounted on a Cu clip, placed into an atom probe puck and loaded in ambient conditions into Side 2 of the dual puck stage plate. This dual puck approach allows the easy transfer of air sensitive substrates into the PFIB and the removal of sharpened APT samples with minimal transfer stages. The sites of interest were removed from the bulk material using $Xe^+$ plasma of the Helios 5 CX Hydra and similar PFIB conditions as in reference.[47] Once the milled-out region was mounted and welded to the top of the post, the sample was sharpened to an apex diameter < 100 nm using 30 kV $Xe^+$ from 1 to 30 pA, and polishing using 5 kV $Xe^+$ ions at 30 pA to remove regions severely damaged by the higher energy ions as in ref. [48]. The samples were then transferred in UHV to the LEAP via the Ferrovac suitcase and analysed in a Cameca LEAP 5000 XR (Local Electrode Atom Probe equipped with reflection) using laser pulsing conditions of 30 pJ, 140-200 kHz and 50 K base temperature. APT data reconstruction and analysis were performed using CAMECA Integrated Visualization and Analysis Software (IVAS) as part of AP Suite 6.

<u>Manuscript</u>

A large language model, ChatGPT, was used for proof reading of the manuscript text.

**Acknowledgements**

The authors thank Dr. Anders Eklund for providing feedback on the final manuscript.

D.M.E. acknowledges and thanks funding from the Deutsche Forschungsgemeinschaft (grant No. EV 305/1-1). M.C. acknowledges funding from the Royal Society Tata University Research Fellowship (URF\R1\201318) and the EPSRC NAME Programme Grant EP/V001914/1. M.C., J.O.D, B.G., are grateful for funding from the EPSRC under the grant EP/V007661/1. D.C., L.Pr. and V.T. acknowledge the support by the Deutsche Forschungsgemeinschaft (DFG, German Research Foundation) TRR 360-492547816 and via Project No. ANCD 20.80009.5007.19 (Moldova). L.Pu. acknowledges the financial support of the DFG via the Transregional Research Collaboration TRR80 (Augsburg, Munich and Stuttgart; Project No. 107745057).

**References**

# References

1. Keimer, B., Kivelson, S. A., Norman, M. R., Uchida, S. & Zaanen, J. From quantum matter to high-temperature superconductivity in copper oxides. *Nature* **518**, 179–86 (2015).

2. MacManus-Driscoll, J. L. & Wimbush, S. C. Processing and application of high-temperature superconducting coated conductors. *Nat. Rev. Mater.* **6**, 587–604 (2021).

3. Fernandes, R. M. *et al.* Iron pnictides and chalcogenides: a new paradigm for superconductivity. *Nature* **601**, 35–44 (2022).

4. Bohmer, A. E. & Kreisel, A. Nematicity, magnetism and superconductivity in FeSe. *J. Phys. Condens. Matter* **30**, 023001 (2018).

5. Zhou, X. *et al.* High-temperature superconductivity. *Nat. Rev. Phys.* **3**, 462–465 (2021).

6. Miccoli, I., Edler, F., Pfnur, H. & Tegenkamp, C. The 100th anniversary of the four-point probe technique: the role of probe geometries in isotropic and anisotropic systems. *J. Phys. Condens. Matter* **27**, 223201 (2015).

7. Campbell, A. M., Blunt, F. J., Johnson, J. D. & Freeman, P. A. Quantitative determination of percentage superconductor in a new compound. *Cryogenics* **31**, 732–737 (1991).

8. Krzton-Maziopa, A., Svitlyk, V., Pomjakushina, E., Puzniak, R. & Conder, K. Superconductivity in alkali metal intercalated iron selenides. *J. Phys. Condens. Matter* **28**, 293002 (2016).

9. Dudin, P. *et al.* Imaging the local electronic and magnetic properties of intrinsically phase separated $Rb_xFe_{2-y}Se_2$ superconductor using scanning microscopy techniques. *Supercond. Sci. Technol.* **32**, 44005 (2019).

10. Eley, S., Gopalakrishnan, S., Goldbart, P. M. & Mason, N. Approaching zero-temperature metallic states in mesoscopic superconductor–normal–superconductor arrays. *Nat. Phys.* **8**, 59–62 (2012).

11. Eley, S., Gopalakrishnan, S., Goldbart, P. M. & Mason, N. Dependence of global superconductivity on inter-island coupling in arrays of long SNS junctions. *J. Phys. Condens. Matter* **25**, 445701 (2013).

12. van der Pauw, L. J. A method of measuring specific resistivity and Hall effect of discs of arbitrary shape. *Philips Res. Rep.* **13**, 1–9 (1958).

13. Li, Y. *et al.* Electronic properties of the bulk and surface states of $Fe_{1+y}Te_{1-x}Se_x$. *Nat. Mater.* **20**, 1221–1227 (2021).

14. Blonder, G. E., Tinkham, M. & Klapwijk, T. M. Transition from metallic to tunneling regimes in superconducting microconstrictions: Excess current, charge imbalance, and supercurrent conversion. *Phys. Rev. B* **25**, 4515–4532 (1982).

15. Daghero, D. & Gonnelli, R. S. Probing multiband superconductivity by point-contact spectroscopy. *Supercond. Sci. Technol.* **23**, 43001 (2010).

16. Naidyuk, Yu. G. & Gloos, K. Anatomy of point-contact Andreev reflection spectroscopy from the experimental point of view. *Low Temp. Phys.* **44**, 257–268 (2018).

17. Murrell, M. P. *et al.* Spatially resolved electrical measurements of $SiO_2$ gate oxides using atomic force microscopy. *Appl. Phys. Lett.* **62**, 786–788 (1993).

18. Puntigam, L. *et al.* Strain Driven Conducting Domain Walls in a Mott Insulator. *Adv. Electron. Mater.* **8**, (2022).

19. Ghara, S. *et al.* Giant conductivity of mobile non-oxide domain walls. *Nat. Commun.* **12**, 3975 (2021).

20. Kalinin, S. V., Borisevich, A. & Fong, D. Beyond condensed matter physics on the nanoscale: the role of ionic and electrochemical phenomena in the physical functionalities of oxide materials. *ACS Nano* **6**, 10423–37 (2012).

21. Evans, Donald. M. *et al.* Conductivity control via minimally invasive anti-Frenkel defects in a functional oxide. *Nat. Mater.* **19**, 1195–1200 (2020).

22. Evans, D. M. *et al.* Observation of Electric-Field-Induced Structural Dislocations in a Ferroelectric Oxide. *Nano Lett.* **21**, 3386–3392 (2021).

23. Kawasegi, N. *et al.* Nanomachining of Silicon Surface Using Atomic Force Microscope With Diamond Tip. *J. Manuf. Sci. Eng.* **128**, 723–729 (2005).

24. Kreisel, A., Hirschfeld, P. & Andersen, B. On the Remarkable Superconductivity of FeSe and Its Close Cousins. *Symmetry* **12**, (2020).

25. Shibauchi, T., Hanaguri, T. & Matsuda, Y. Exotic Superconducting States in FeSe-based Materials. *J. Phys. Soc. Jpn.* **89**, (2020).

26. Tsurkan, V. *et al.* Physical properties of FeSe$_{0.5}$Te$_{0.5}$ single crystals grown under different conditions. *Eur. Phys. J. B* **79**, 289–299 (2011).

27. Homes, C. C., Dai, Y. M., Wen, J. S., Xu, Z. J. & Gu, G. D. FeTe$_{0.55}$Se$_{0.45}$: A multiband superconductor in the clean and dirty limit. *Phys. Rev. B* **91**, 144503 (2015).

28. Cho, K. *et al.* Precision global measurements of London penetration depth in FeTe$_{0.58}$Se$_{0.42}$. *Phys. Rev. B* **84**, 174502 (2011).

29. Cho, D., Bastiaans, K. M., Chatzopoulos, D., Gu, G. D. & Allan, M. P. A strongly inhomogeneous superfluid in an iron-based superconductor. *Nature* **571**, 541–545 (2019).

30. Chatzopoulos, D. *et al.* Spatially dispersing Yu-Shiba-Rusinov states in the unconventional superconductor FeTe$_{0.55}$Se$_{0.45}$. *Nat. Commun.* **12**, 298 (2021).

31. Liu, C. *et al.* Zero-energy bound states in the high-temperature superconductors at the two-dimensional limit. *Sci. Adv.* **6**, eaax7547 (2020).

32. Machida, T. *et al.* Zero-energy vortex bound state in the superconducting topological surface state of Fe(Se,Te). *Nat. Mater.* **18**, 811–815 (2019).

33. Yin, J. X. *et al.* Observation of a robust zero-energy bound state in iron-based superconductor Fe(Te,Se). *Nat. Phys.* **11**, 543–546 (2015).

34. Wang, Z. *et al.* Topological nature of the FeTe$_{0.5}$Se$_{0.5}$ superconductor. *Phys. Rev. B* **92**, 115119 (2015).

35. Wang, Z. *et al.* Evidence for dispersing 1D Majorana channels in an iron-based superconductor. *Science* **367**, 104–108 (2020).

36. Singh, U. R. *et al.* Spatial inhomogeneity of the superconducting gap and order parameter in FeSe$_{0.4}$Te$_{0.6}$. *Phys. Rev. B* **88**, 155124 (2013).

37. Hanaguri, T. *et al.* Quantum Vortex Core and Missing Pseudogap in the Multiband BCS-BEC Crossover Superconductor FeSe. *Phys. Rev. Lett.* **122**, 077001 (2019).

38. Chen, Q., Wang, Z., Boyack, R., Yang, S. & Levin, K. When superconductivity crosses over: From BCS to BEC. *Rev. Mod. Phys.* **96**, 025002 (2024).

39. Sun, Y., Shi, Z. & Tamegai, T. Review of annealing effects and superconductivity in $Fe_{1+y}Te_{1-x}Se_x$ superconductors. *Supercond. Sci. Technol.* **32**, 103001 (2019).

40. Bean, C. P. Magnetization of Hard Superconductors. *Phys. Rev. Lett.* **8**, 250–253 (1962).

41. Bean, C. P. Magnetization of High-Field Superconductors. *Rev. Mod. Phys.* **36**, 31–39 (1964).

42. Rinott, S. *et al.* Tuning across the BCS-BEC crossover in the multiband superconductor $Fe_{1+y}Se_xTe_{1-x}$: An angle-resolved photoemission study. *Sci. Adv.* **3**, e1602372 (2017).

43. Gault, B. *et al.* Atom probe tomography. *Nat. Rev. Methods Primer* **1**, 51 (2021).

44. Singh, U. R. *et al.* Evidence for orbital order and its relation to superconductivity in $FeSe_{0.4}Te_{0.6}$. *Sci. Adv.* **1**, e1500206 (2015).

45. Wahl, P., Singh, U. R., Tsurkan, V. & Loidl, A. Nanoscale electronic inhomogeneity in $FeSe_{0.4}Te_{0.6}$ revealed through unsupervised machine learning. *Phys. Rev. B* **101**, 115112 (2020).

46. Nečas, D. & Klapetek, P. Gwyddion: An open-source software for SPM data analysis. *Cent. Eur. J. Phys.* **10**, 181–188 (2012).

47. Thompson, K. *et al.* In situ site-specific specimen preparation for atom probe tomography. *Ultramicroscopy* **107**, 131–139 (2007).

48. Douglas, J. O., Conroy, M., Giuliani, F. & Gault, B. In Situ Sputtering From the Micromanipulator to Enable Cryogenic Preparation of Specimens for Atom Probe Tomography by Focused-Ion Beam. *Microsc. Microanal.* **29**, 1009–1017 (2023).

49. https://www.attocube.com/en/products/microscopes/scanning-probe-microscopes/attoafmi-1. *Attocube website*.

50. https://www.zhinst.com/europe/en/applications/scanning-probe-microscopy/scanning-near-field-microscopy-snom.

51. Pan, C., Shi, Y., Hui, F., Grustan-Gutierrez, E. & Lanza, M. History and Status of the CAFM. in *Conductive Atomic Force Microscopy* 1–28 (2017). doi:10.1002/9783527699773.ch1.

52. Benstetter, G., Hofer, A., Liu, D., Frammelsberger, W. & Lanza, M. Fundamentals of CAFM Operation Modes. in *Conductive Atomic Force Microscopy* 45–77 (2017). doi:10.1002/9783527699773.ch3.

53. Mele, E. J. Screening of a point charge by an anisotropic medium: Anamorphoses in the method of images. *Am. J. Phys.* **69**, 557–562 (2001).

54. Wang, B. & Woo, C. H. Atomic force microscopy-induced electric field in ferroelectric thin films. *J. Appl. Phys.* **94**, 4053–4059 (2003).

55. Abbas, Y. *et al.* Photodetection Characteristics of Gold Coated AFM Tips and n-Silicon Substrate nano-Schottky Interfaces. *Sci. Rep.* **9**, 13586 (2019).

56. Fischer, Ø., Kugler, M., Maggio-Aprile, I., Berthod, C. & Renner, C. Scanning tunneling spectroscopy of high-temperature superconductors. *Rev. Mod. Phys.* **79**, 353–419 (2007).

57. Singh, U. R. *et al.* Spatial inhomogeneity of the superconducting gap and order parameter in FeSe0.4Te0.6. *Phys. Rev. B* **88**, 155124, publisher = American Physical Society (2013).

58. Iranmanesh, M., Stir, M., Kirtley, J. R. & Hulliger, J. Scanning SQUID Microscopy of Local Superconductivity in Inhomogeneous Combinatorial Ceramics. *Chem. – Eur. J.* **20**, 15816–15823 (2014).

59. Kirtley, J. R. Fundamental studies of superconductors using scanning magnetic imaging. *Rep. Prog. Phys.* **73**, 126501 (2010).

60. Dutta, S. K. *et al.* Imaging microwave electric fields using a near-field scanning microwave microscope. *Appl. Phys. Lett.* **74**, 156–158 (1999).

61. Sobota, J. A., He, Y. & Shen, Z.-X. Angle-resolved photoemission studies of quantum materials. *Rev. Mod. Phys.* **93**, 25006 (2021).

**Supplementary Information**

**Supplementary Note 1 | Introduction to cAFM**

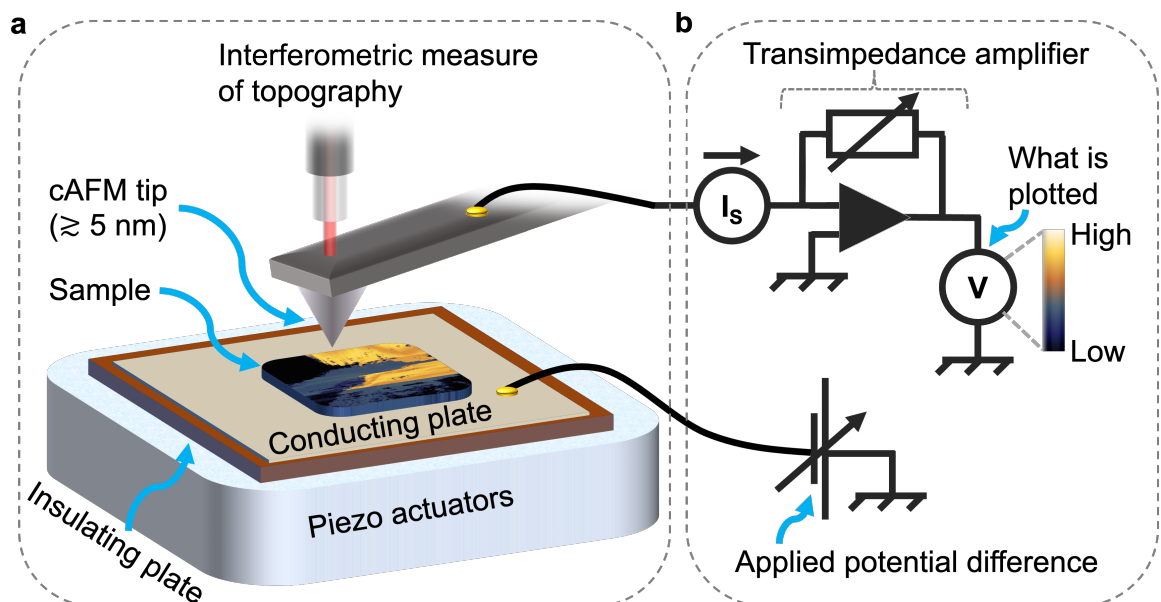

**Figure S1 | Schematic of cAFM setup as used in an attocube AFM 1. a,** An optical interferometer detects cantilever deflection; the feedback loop adjusts the Z-piezo to hold the interference signal from the reflection constant. This aims to ensures a constant tip-sample force even when scanning rough topographic features. **b,** Current is measured independently using a low-noise transimpedance amplifier, which converts the tip–sample current into a voltage signal. Variations in conductivity are displayed using false colour throughout this manuscript. Topography and conductivity signals are acquired simultaneously and independently.

Conductive atomic force microscopy (cAFM) is a scanning-probe technique in which a sharp, conductive tip is rastered across the sample surface. A feedback loop adjusts the extension of the piezo actuators to minimise deviations of the cantilever deflection from a defined set point, thereby maintaining a constant normal force. The cantilever deflection and the height signal from the piezo actuators are recorded continuously and saved as the intensity and topography data, respectively.

In the contact-mode implementation used here, once the tip is in contact with the sample, the instrument scans each line forward and then backward along the fast-scan axis before stepping to the next line. Under stable conditions, the topography from the two traces is expected to be identical. CAFM therefore yields three independent data channels per pixel—surface topography, cantilever deflection, and local electrical current—collected by separate detectors (schematic in Figure S1 and details on the manufacturer's webpage).[49]

We used commercially available, boron-doped diamond tips with apex radii < 10 nm; their high conductivity, mechanical robustness, and nanometric contact area provide long-term stability and high spatial resolution.[50] These cAFM tips are orders of magnitude sharper than the tens-of-micrometre end diameters commonly used in PCS measurements.[15] This implies that, compared to PCS, c-AFM measurements are typically performed in the ballistic regime.[15]

A bias of typically ±(1 to 10) V is applied between the tip and the bottom electrode, driving a current through the tip–sample junction. The analogue current is converted to a digital signal by a low-noise current trans-impedance amplifier (see schematic in Figure S1).[51] The measured current, $I$, equals the product of the local current density, $J$, and the effective emission area $A_{eff}$:

$$I = J \times A_{eff}.$$

$A_{eff}$ is distinct to the nanometric mechanical contact area, as the electric field may spread laterally through high-conductivity regions, and may encompass the entire footprint of a good metallic pad [51,52] In the case of thick dielectric films, the contact is well approximated by Hertzian mechanics, and the electric field decays within a few tip radii.[53–55] The total junction resistance includes (i) geometric spreading resistance and (ii) interface terms that depend on the work-function mismatch between tip and sample; reported offsets range from –0.3 V to +1.5 V for common tips on semiconductors.[52] Even an ostensibly ideal Pt-on-Pt junction the best resistance reported was ≈ 300 Ω.[52] Together with the sheet resistance inherent to the thin metallic coating on cAFM tips, and additional interfacial artefacts, this limits the absolute accuracy of conductivity values extracted from c-AFM. Consequently, the technique is most reliable for relative comparisons, and quantitative analysis requires meticulous calibration and modelling.

**Supplementary Note 2 | Evaluating conductivity in superconductors**
To the best of our knowledge, there is no established method for directly measuring and spatially resolving local variations in DC conductivity in superconductors. There are several techniques where the measured signal is evaluated to deduce conductivity: (i) scanning superconducting quantum interference devices (S-SQUIDs), which detect stray magnetic fields; (ii) scanning microwave impedance microscopy (sMIM), which probes local permittivity and conductivity via reflected microwave fields; (iii) angle-resolved photoemission spectroscopy (ARPES), which measures the momentum-resolved electronic band structure; and (iv) scanning tunnelling microscopy (STM), which probes the local surface density of states (sDOS).[42,56–61] However, none of these techniques directly measures the local variations in DC conductivity—the critical quantity required to validate spatially resolved four-probe transport measurements.

**Supplementary Note 3 | Evaluating superconducting volume traction**

Considering the simple Debye approximation for specific heat, $C = \gamma T + \beta T^3$, where $\gamma$ is the Sommerfeld coefficient, related to the electronic contributions, and $\beta$ represents the contribution from the lattice we can replot our specific heat data in the form $C/T$ vs $T^2$. Below 5 K where the simple Debye plot is considered reasonable, a linear fit gives values of $\gamma_0$ = 0.807 mJ / mol K$^2$ and $\beta$ = 0.393. Note that in this data there appears to be an additional contribution appears below 3 K, which is believed to originate from a measurement artifact.

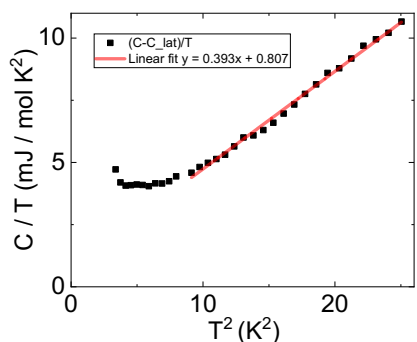

**Figure S2 | Temperature dependence of specific heat.** The y-axis intercept, 0.807 mJ/mol K$^2$, gives the value of $\gamma_0$ in a simplistic Debye situation.

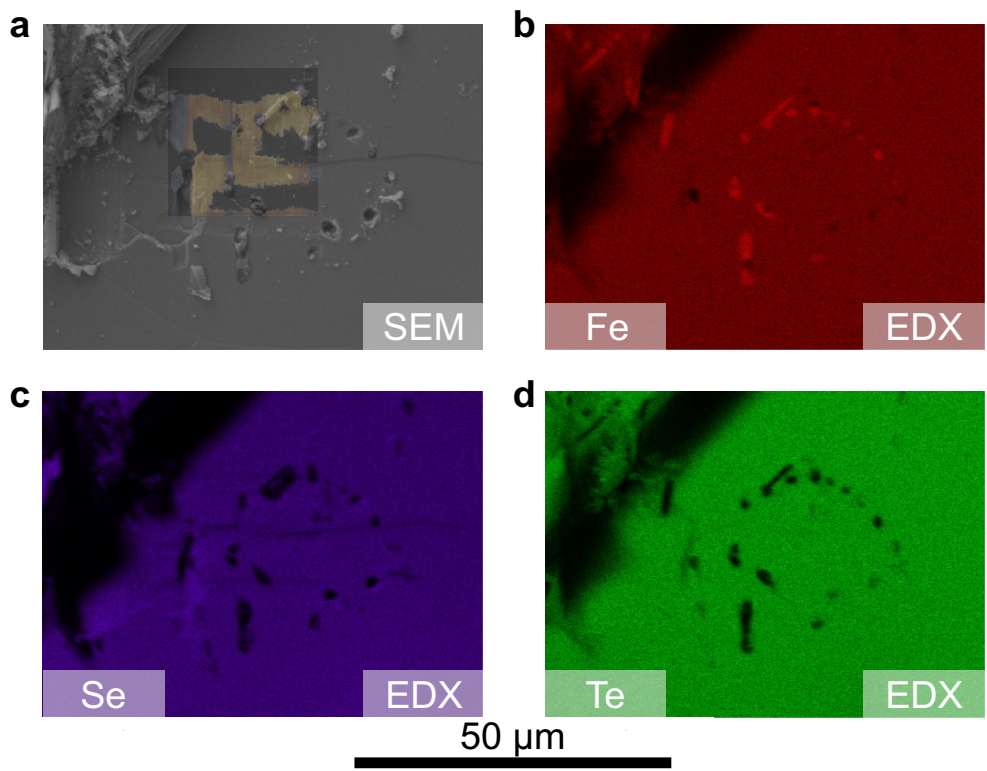

**Figure S3 | Heterogeneous conductivity area in EDX. a,** A superposition of the cAFM data onto a large SEM image, the images were aligned using the topography data collected during the cAFM scan. This is the same area discussed in Figures 3, 4, and 5 of the manuscript. **b, c,** and **d,** EDX map of iron, selenium and tellurium content respectively. All images show an apparently homogeneous crystal with iron rich detritus on the surface.

Critically this shows EDX on its own cannot identify the compositional variations associated with the enhanced conductivity, even when mapping the specific region.

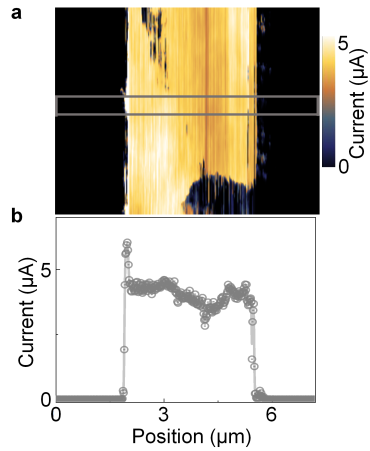

**Figure S4 | Abrupt transition between superconducting and normal regions. a**, Cropped data set of the region plotted in Figure 3, the grey box indicates the area averaged for the cross section. **b**, Graphical representation of the cross-section area marked in **a**. The transition between no measurable current and very large currents is abrupt happening across ca. 50 nm.

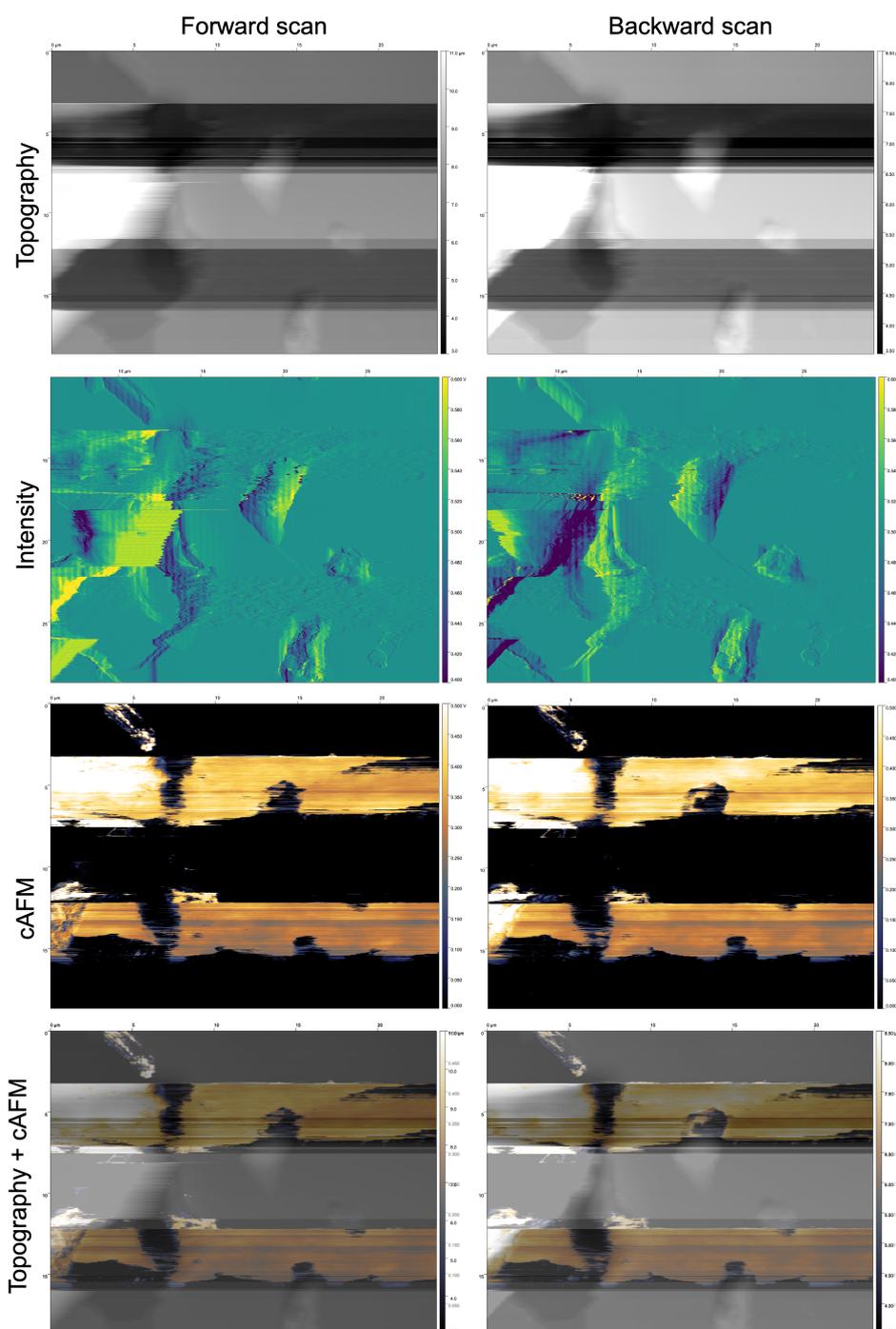

**Figure S5 | Topography, intensity, and cAFM data for forward and backward scans on the same region (fast scan axis parallel to stripes).** Despite minor line-to-line jumps in topography, these jumps are multiple pixels away from the changes in conductivity, meaning that the *changes* in topography does not affect the cAFM. The intensity shows the deflection of the cantilever by the sample, in this experiment we worked on a falling flank and contact was defined as 0.5 V: so numbers below 0.5 V (darker colours) indicated an increase in tip-sample pressure. Conductivity variations occur both with and without changes in height, and do not correlate with the significant changes in pressure from the intensity images. This indicates they are not imaging artifacts, and the areas with no conductivity have good tip-sample contact and nice topographic data. Data was acquired with 17 nm pixel spacing and 6 ms/pixel dwell time. None of the standard gwyddion post-processing steps was applied to these images, so there has been no background subtraction, no stroke correction, and no line alignment.[46]